\documentclass[twoside,slac_one]{revtex4}
\usepackage{graphicx}
\usepackage{fancyhdr}
\usepackage{amsmath} 
\usepackage{bm}
\usepackage{amsxtra}
\usepackage{amssymb}
\usepackage{amsthm}
\usepackage{latexsym}
\usepackage{lscape}

\newcommand {\plainafb}	{\ensuremath{A_{\mathrm{FB}}}}
\newcommand {\afb}	{{\plainafb}}

\newcommand {\pbar}	{{\ensuremath{\bar p}}}
\newcommand {\ppbar}	{{\ensuremath{p\pbar}}}
\newcommand {\tbar}     {{\ensuremath{\bar t}}}
\newcommand {\ttbar}    {{\ensuremath{t\tbar}}}

\newcommand {\pjets}    {{\textrm{+jets}}}
\newcommand {\wpj}      {{\ensuremath{W\!+}jets}}
\newcommand {\wpjmath}  {{W\!+\mathrm{jets}}}

\newcommand {\lpj}      {{\ensuremath{l\pjets}}}
\newcommand {\epj}      {{\ensuremath{e\pjets}}}
\newcommand {\mpj}      {{\ensuremath{\mu\pjets}}}

\newcommand {\GeV}        {{\ensuremath{\,\textrm {GeV}}}}

\newcommand {\ifb}        {{\ensuremath{\,\textrm {fb}^{-1}}}}

\newcommand{\lep}         {\ensuremath{l}} 
\newcommand {\pt}         {{\ensuremath{p_T}}}
\newcommand {\ylep}       {{\ensuremath{y_\lep}}}
\newcommand {\qlep}       {{\ensuremath{q_\lep}}}
\newcommand {\qyl}        {{\ensuremath{\qlep\ylep}}}

\newcommand {\dy}         {{\ensuremath{\Delta y}}}
\newcommand {\absdy}      {{\ensuremath{\left|\dy\right|}}}

\newcommand {\ttpt} {{\ensuremath{p_T^\ttbar}}}

\newcommand{\emiss} {{/\!\!\!\!E}}
\newcommand{\met}   {{\ensuremath{\emiss_T}}}

\newcommand {\afbl}   {{\ensuremath{\plainafb^\lep}}}
\newcommand {\mttbar} {{\ensuremath{m_\ttbar}}}

\newcommand {\Ntt}    {{\ensuremath{N_{\ttbar}}}}
\newcommand {\Nw}     {{\ensuremath{N_{\wpjmath}}}}
\newcommand {\Nmj}    {{\ensuremath{N_{\mathrm{MJ}}}}}
\newcommand {\Nf}     {{\ensuremath{N_F}}}
\newcommand {\Nb}     {{\ensuremath{N_B}}}
\newcommand {\Nfl}    {{\ensuremath{N_F^\lep}}}
\newcommand {\Nbl}    {{\ensuremath{N_B^\lep}}}
\newcommand {\Nfwd}   {\Nf}
\newcommand {\Nbwd}   {\Nb}
\newcommand {\Nlfwd}  {\Nfl}
\newcommand {\Nlbwd}  {\Nbl}

\newcommand {\DZ}     {{D0}} 

 \newcommand {\pythia}   {{\sc pythia}}
 \newcommand {\alpgen}   {{\sc alpgen}}
 \newcommand {\mcatnlo}  {{\sc mc@nlo}}
 \newcommand {\herwig}   {{\sc herwig}}

 \newcommand {\ie}       {{\rm i.e.}}
 
 \newcommand {\etal}     {{\it et al.}}

\begin{document}

\title{Measurement of the forward-backward asymmetry in top-antitop quark production in proton-antiproton collisions}

%

\author{D. Orbaker for the D\O~collaboration}
\affiliation{Department of Physics and Astronomy, University of Rochester, Rochester, NY, USA}

\begin{abstract}

We present a new measurement of the forward-backward asymmetry in \ttbar\ production in \ppbar\ collisions in the \lpj\ channel at \DZ. We measure asymmetries from two different observables and unfold the data to allow the results to be compared to standard model predictions. For the unfolded asymmetry based on the $\dy$ variable, we measure (19.6 $\pm$ 6.5)\%, compared with an \mcatnlo-based prediction of $(5.0 \pm 0.1)\%$. We also discuss the correlation between the asymmetry and the transverse momentum of the \ttbar\ system.

\end{abstract}

\maketitle

\thispagestyle{fancy}


\section{Introduction}

In this note we present a new measurement of the forward-backward asymmetry (\afb) in \ttbar\ production in \ppbar\ collisions using the \lpj\ channel at \DZ~\cite{Abazov:2011rq}. Using 5.4 \ifb\ of data, this measurement is an update to the first \DZ\ result using 0.9 \ifb~\cite{2007qb}. Recently, the CDF collaboration released a result presenting an asymmetry that is a 2-3 sigma deviation from the inclusive standard model (SM) prediction and has a dependence on the invariant mass of the \ttbar\ system, \mttbar~\cite{Aaltonen:2011kc}.

In the SM, the main contribution to \afb\ in \ttbar\ events comes from quantum chromodynamics (QCD). The first calculation for asymmetry at the Tevatron, made more than a decade ago, predicts an \afb\ of $\sim5\%$~\cite{Kuhn:1998jr}. More recent calculations, some of which include electroweak (EW) effects, predict asymmetries up to $\sim9\%$~\cite{Kidonakis:2011zn, Ahrens:2011uf, Hollik:2011ps}. Recent theoretical efforts have been made to include the asymmetry in a framework of measurements including top quark polarization and spin correlation~\cite{Krohn:2011tw}.

To compare our results with theory we present two different types of results. The first type are reconstruction level results. Reconstruction level asymmetries appear in the \DZ\ detector, after the effects of detector acceptance and event reconstruction. Quantities measured at the reconstruction level cannot be directly compared with theory. For this reason we also present the measurement at the production level, before effects of selection and reconstruction. To access the production level for a given quantity we unfold the distributions in data after background subtraction.

\section{The \DZ\ Detector}

\DZ\ is a general purpose particle detector~\cite{run2det}. Going outward from the beam pipe, the first subsystem of the detector is the central-tracking system, which consists of a silicon microstrip tracker (SMT), as well as a central fiber tracker (CFT). Both portions of the tracker are located inside a 2~T superconducting solenoid magnet. The designs are optimized to track and vertex charged particles at rapidities up to $|\eta|<3$ for the SMT and $|\eta|<2.5$ for the CFT. Right outside of the solenoid are central and forward preshower detectors. Three liquid argon and uranium calorimeters sit outside of the preshower: one central section (CC), which covers $|\eta|$ up to $\approx 1.1$ and two end sections (EC) that extend the range in $|\eta|$ to $\approx 4.2$. All three sections of the calorimeter are housed in separate cryostats~\cite{run1det}. Furthest away from the collision region is the muon system, which covers a range of $|\eta|<2$ and consists of a layer of tracking detectors and scintillator trigger counters, followed by 1.8~T toroids and then two similar detector and scintillator layers~\cite{run2muon}. Plastic scintillator arrays placed in front of the EC cryostats measure the luminosity. Both the trigger and data acquisition systems are designed to handle the high instantaneous luminosities of the Tevatron Run II.

\section{Event Selection and Reconstruction}

The requirements for events to be selected are very similar to those found in the most recent \DZ\ \lpj\ crosssection measurement~\cite{Abazov:2011mi}. In the \lpj\ channel, the top quark pair decays as such: $\ttbar\to W^{+}bW^{-}\bar{b}$, with one $W$ boson decaying leptonically, $W\to l\nu$ and the other $W$ boson decaying hadronically $W\to q \bar{q}^{'}$. Looking at $\ttbar \to b\bar{b}l\nu q\bar{q}^{'}$, one can see that there are six decay products in the final state: a lepton, which in this measurement is either a muon ($\mu$) or an electron ($e$), missing transverse energy from the neutrino (\met) and four jets, two of them jets originating from $b$ quarks. 

Keeping the six decay products in mind, we require four jets with transverse momentum (\pt) greater than 20 GeV and pseudorapidity ($|\eta|$) less than 2.5. At least one of these jets must $\pt > 40$GeV. We require one and only one lepton; either an electron with $\pt > 20$GeV and $|\eta| < 1.1$ or a muon with $\pt > 20$GeV and $|\eta| < 2.0$. Missing transverse energy must be greater than 20 GeV in the \epj\ channel and 25 GeV in the \mpj\ channel.  

Jets originating from $b$ quarks are identified by displaced vertices via a $b$-tagging neural net algorithm~\cite{Abazov:2010ab}. Every event must have at least one jet that passes the $b$-tagging requirements. The efficiency for a $b$ jet to pass the $b$-tagging requirements is $\sim$70\%, while likelihood of a jet originating from quarks of lighter weight to mimic a $b$-jet and be misidentified is $\sim$8\%. For interested readers, more detail may be found here~\cite{Abazov:2007kg}.

We employ a constrained kinematic fit to reconstruct the six decay objects into the top quark four vectors under the \ttbar\ decay hypothesis~\cite{bib:hitfit}. A $\chi^{2}$ function based on the detector resolution is minimized for each jet-parton permutation and only the assignment with the lowest $\chi^{2}$ is kept. To reduce the number of possible assignments, at least one of the jets associated with a $b$ quark by the kinematic fit must be $b$-tagged. A total of four constraints are used: each reconstructed top quark must have a mass of 172.5 GeV and each reconstructed $W$ boson must have a mass of 80.4 GeV. The jet-parton assignment is correct $\sim70\%$ of the time.

\section{Simulation and Backgrounds}

Events from \ttbar\ decays are simulated with \mcatnlo, with \herwig\ used for showering~\cite{bib:mcatnlo, bib:herwig}. Background events from $W$ boson produced in association with jets ($W$+jets) decays are simulated in a similar fashion with \alpgen+\pythia~\cite{bib:alpgen, bib:pythia}. We also take into account background events from multijet (MJ) production, where one jet mimics the signature of a lepton.  Other backgrounds such as single top, diboson and $Z$+jets are insignificant and not considered in this analysis. To be comparable to data, events from the Monte Carlo simulations are run through the \DZ\ detector simulation~\cite{bib:run2det} and the reconstruction sequence used on data. To evaluate the contribution from the MJ background we use control samples from \DZ\ data.

\section{Definition of Asymmetries}

The forward-backward asymmetry is one of the quantities useful for summarizing a differential distribution. The forward-backward asymmetry is the difference between the fraction of events defined as forward ($N_{f}$) and the fraction of events defined as backward ($N_{b}$),

\begin{equation}
\afb = \frac{N_{f} - N_{b}}{N_{f} + N_{b}}.
\end{equation}

We present the asymmetry in two different observables: $\dy\ = y_t - y_{\bar{t}} = q_{l}(y_{t,lep}-y_{t,had})$ and \qyl, where the rapidity is defined as $y = \frac{1}{2} ln\left( \frac{E+p_{z}}{E-p_{z}}\right)$, $q_{l}$ is the charge of the lepton, $t,lep$ is the leptonically decaying top quark and $t,had$ is the hadronically decaying top quark. 

For \dy, the asymmetry is 

\begin{equation}
\afb = \frac{N(\dy > 0) - N(\dy < 0)}{N(\dy > 0) + N(\dy < 0)}.
\end{equation}

The forward-backward asymmetry for \qyl, the so-called lepton-based asymmetry, is

\begin{equation}
\afbl = \frac{N(\qyl > 0) - N(\qyl < 0)}{N(\qyl > 0) + N(\qyl < 0)}.
\end{equation}

\section{Predicted asymmetries}
We predict \afb\ and \afbl\ at the reconstruction level using the events generated from \mcatnlo\ and run through the \DZ\ detector simulation with the same selection criteria and reconstruction used on data. The predicted asymmetries for both the reconstruction and production levels are summarized in Table~\ref{tab:preds}. Production level asymmetries are taken before events are run through through the \DZ\ detector simulation.

\begin{table}[htbp]
\caption{Predictions from the \rm{MC@NLO} event generator.
  \label{tab:preds}
}
\begin{tabular}{|l|lc|c|}
\hline
 & Channel & \afb\ (\%) & \afbl\ (\%) \\
\hline 
Generated& \lpj & $5.0\pm0.1$ & $2.1\pm0.1$ \\[2ex]
Reconstructed 
         & \epj & $2.4\pm0.7$ & $0.7\pm0.6$ \\
         & \mpj & $2.5\pm0.9$ & $1.0\pm0.8$  \\
         & \lpj & $2.4\pm0.7$ & $0.8\pm0.6$ \\
\hline
\end{tabular}
\end{table}

\section{Reconstruction Level Technique and Results}

Both the amount of events from signal and background processes,  \ie\ the sample composition, and the asymmetry from \ttbar\ events are measured via a template-based maximum likelihood (ML) fit. A discriminant made up of four input variables shown in Figure~\ref{fig_disc_inputs} that are loosely correlated with \dy\ is trained to separate the \wpj\ background from the \ttbar\ signal. Four different templates containing the discriminant and the sign of the asymmetry are fit to the data: a signal template made up with events where $\dy > 0$; another signal template made up with events where $\dy < 0$; the template for the \wpj\ background with the asymmetry taken from simulation; the template for the MJ background with the discriminant shape and asymmetry taken from the control data sample.

The results using the method described in the previous paragraph for asymmetries from both \dy\ and \qyl\ are presented in Tables~\ref{tab:afbcomb} and~\ref{tab:afbl}, respectively.  For the lepton-based measurement, an additional selection criterion of $|y_{l}| < 1.5$ is applied. To search for new physics, the asymmetry is also presented for different invariant mass regions and different magnitudes of \dy\ in Table~\ref{tab:subsamples}.

\begin{figure}[ht]
\centering
\includegraphics[width=70mm]{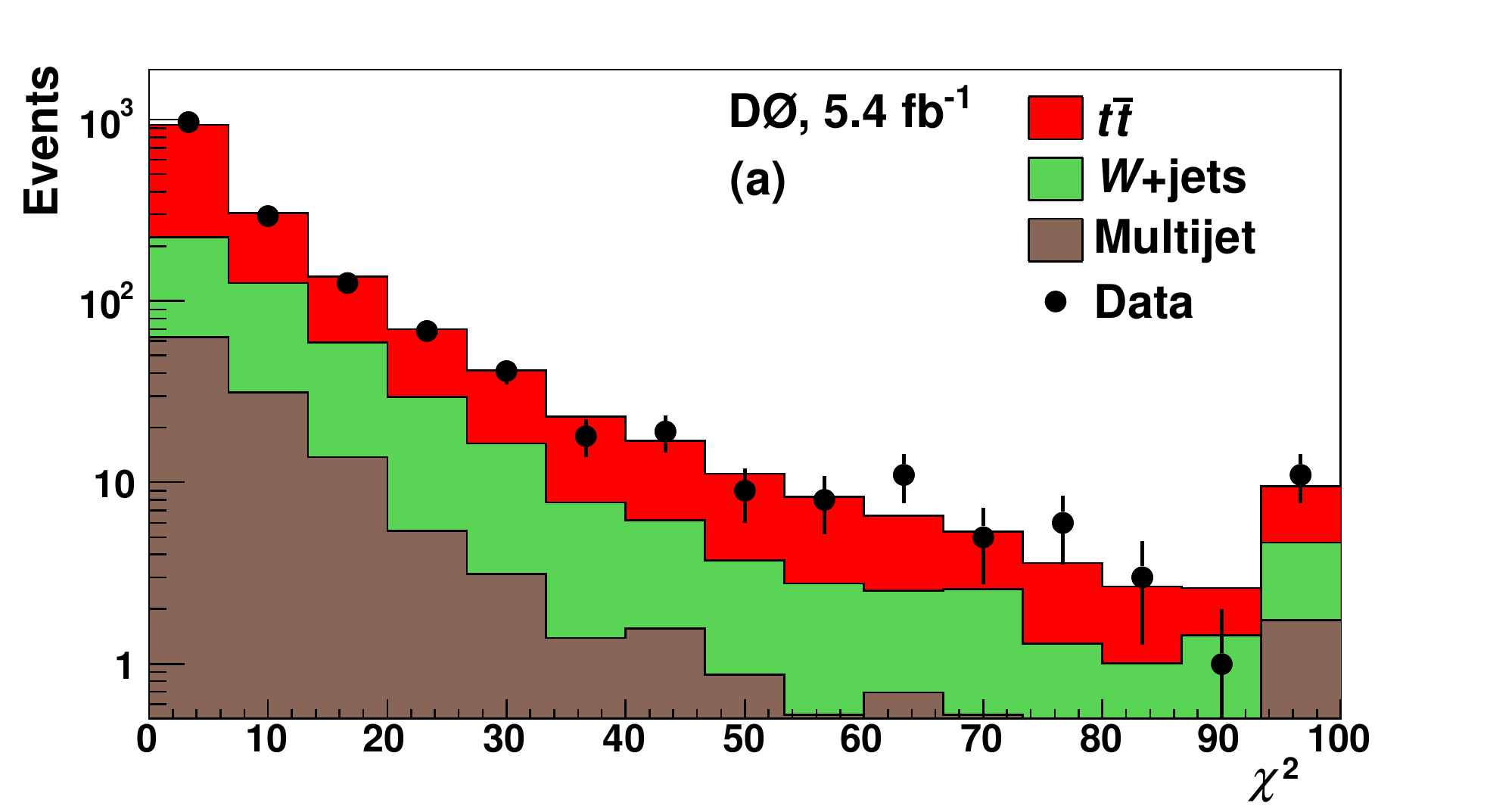}
\includegraphics[width=70mm]{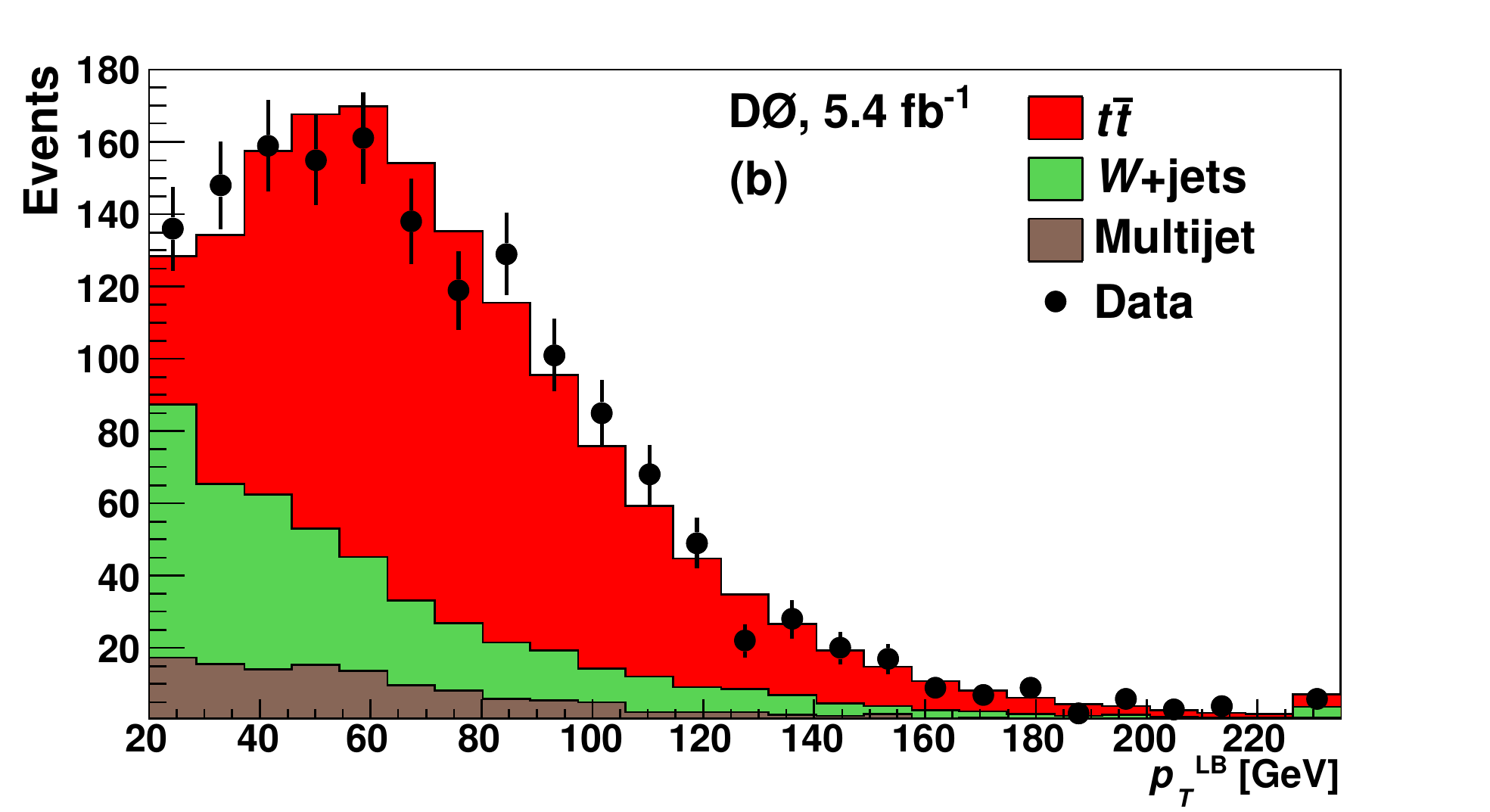}
\includegraphics[width=70mm]{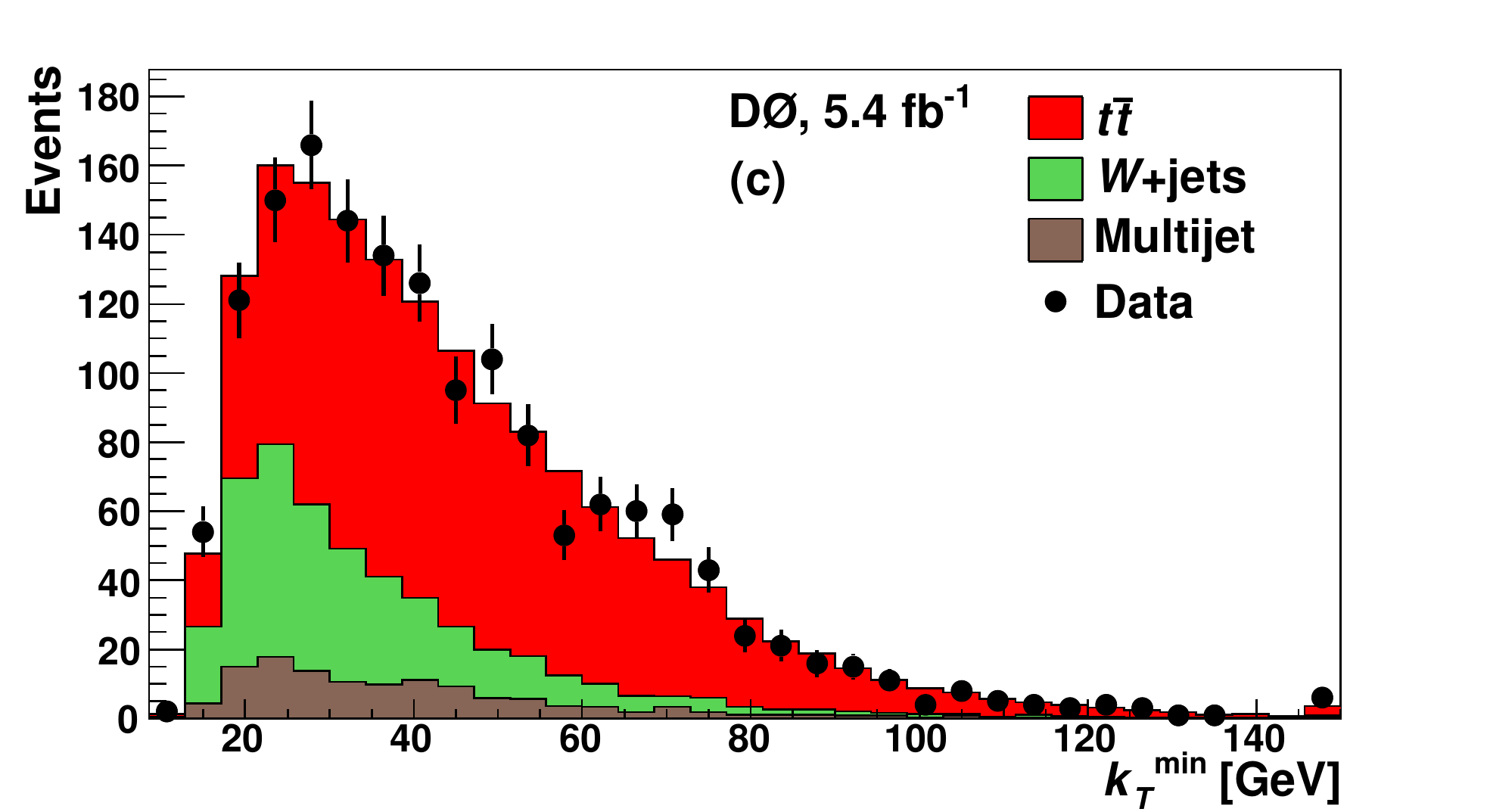}
\includegraphics[width=70mm]{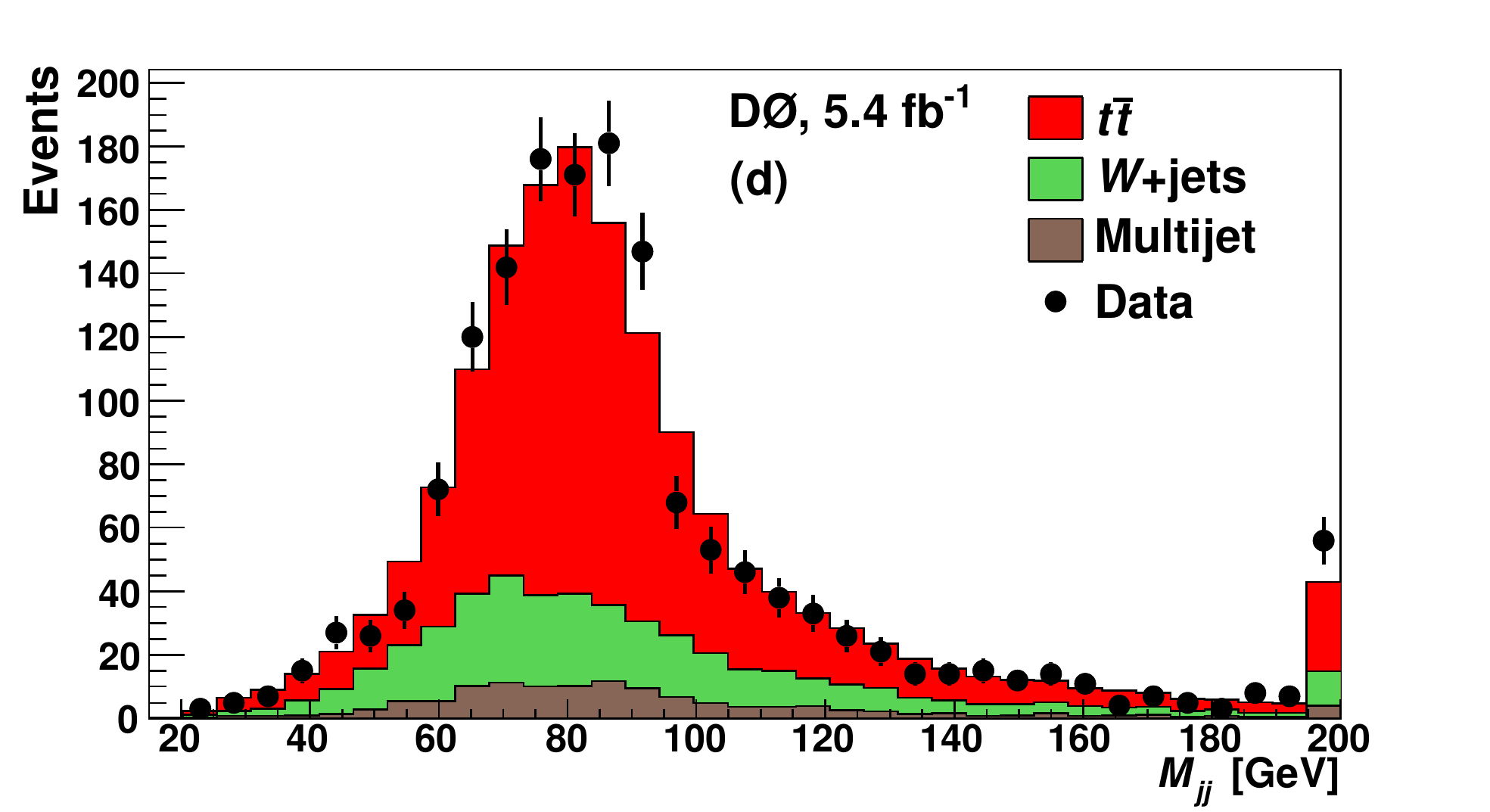}
\caption{Inputs to the discriminant used for the reconstruction level measurement. a) $\chi^{2}$ from the kinematic fit. b) Transverse momentum of the $b$-tagged jet with the largest transverse momentum. c) The minimum relative momentum between any two of the hardest four jets. d) Invariant mass of the two jets assigned by the kinematic fit as daughters of the hadronically decaying $W$ boson.} \label{fig_disc_inputs}
\end{figure}

\begin{figure}[ht]
\centering
\includegraphics[width=70mm]{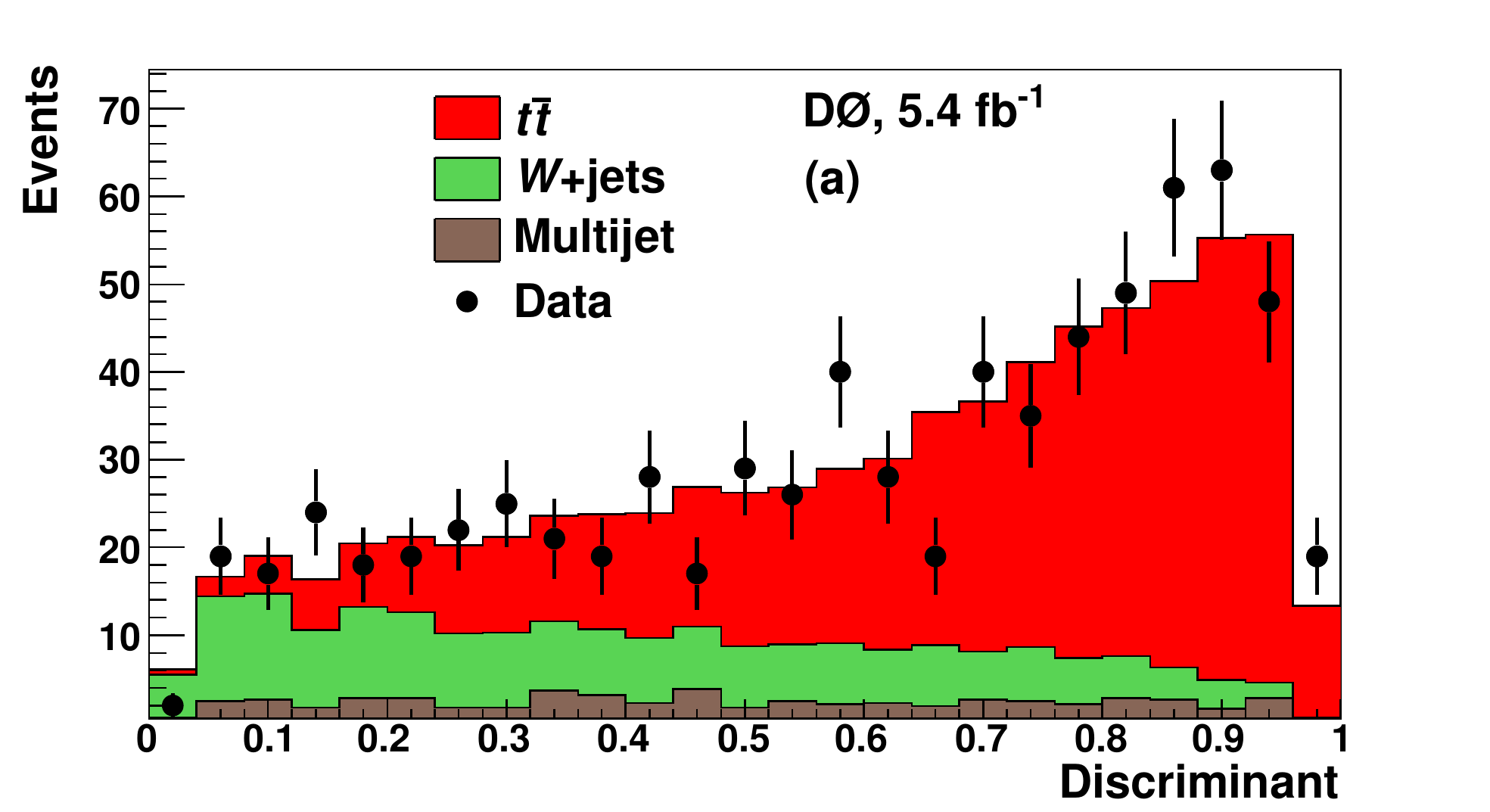}
\includegraphics[width=70mm]{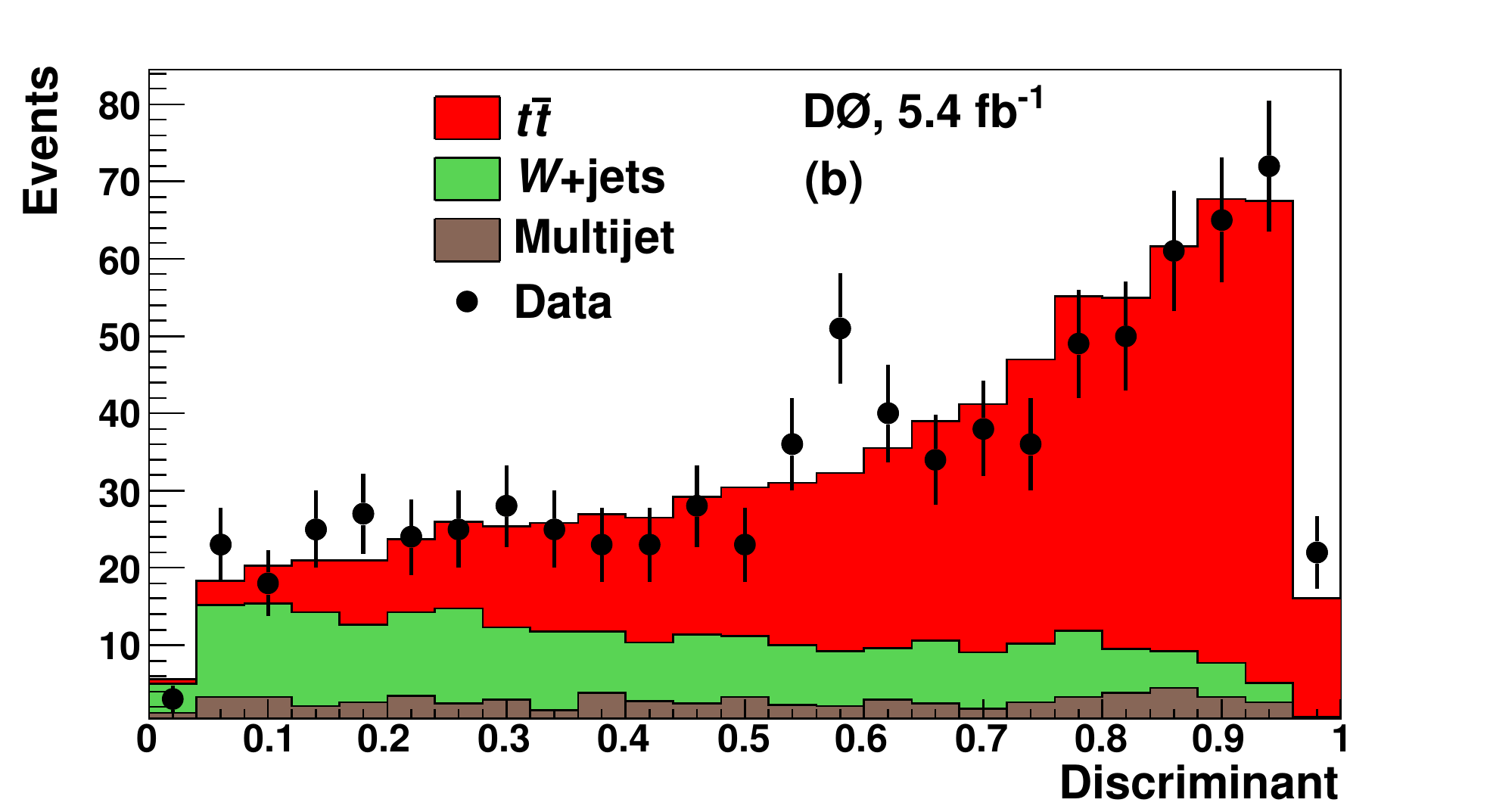}
\caption{The distribution in the discriminant for data and simulation for events with $\dy < 0$ (a) and $\dy > 0$ (b).  \label{fig_disc} }
\end{figure}

\begin{table}[htbp]
\caption{Numbers of events in data, results of fits for sample composition and \afb, 
         and predictions for \afb. The asymmetries are given at reconstruction level,
         with their total uncertainties. The sample compositions are given with 
         the fit uncertainties.
  \label{tab:afbcomb}
}
\begin{tabular}{| l | c | c | c | c | c |}
\hline
    & \multicolumn{1}{|c|}{$l$+$\ge$4 jets} & \multicolumn{1}{|c|}{$e$+$\ge$4 jets} & 
      \multicolumn{1}{|c|}{$\mu$+$\ge$4 jets} & \multicolumn{1}{|c|}{$l$+4 jets} & \multicolumn{1}{|c|}{$l$+$\ge$5 jets} \\
    \hline 
    Raw \Nfwd & \multicolumn{1}{|c|}{849} & \multicolumn{1}{|c|}{455} &
      \multicolumn{1}{|c|}{394} & \multicolumn{1}{|c|}{717} & \multicolumn{1}{|c|}{132}\\
    Raw \Nbwd & \multicolumn{1}{|c|}{732} & \multicolumn{1}{|c|}{397} & \multicolumn{1}{|c|}{335} &
      \multicolumn{1}{|c|}{597} & \multicolumn{1}{|c|}{135}\\[2ex]
    \Ntt       & 1126$\pm$39 & 622$\pm$28 & 502$\pm$28 & 902$\pm$36 & 218$\pm$16\\
    \Nw        &  376$\pm$39 & 173$\pm$28 & 219$\pm$27 & 346$\pm$36 &  35$\pm$16\\
    \Nmj       &   79$\pm$5 &  56$\pm$3 &   8$\pm$2 &  66$\pm$4 &  13$\pm$2\\
    \afb (\%) &    9.2$\pm$3.7 &   8.9$\pm$5.0 &   9.1$\pm$5.8 &  12.2$\pm$4.3 &  -3.0$\pm$7.9\\[2ex]
    \mcatnlo\ \afb\ (\%) &    2.4$\pm$0.7 &  2.4$\pm$0.7   &   2.5$\pm$0.9  &  3.9$\pm$0.8 &  -2.9$\pm$1.1\\
  \hline
  \end{tabular}

\end{table}

\begin{table}[htbp]
\caption{Numbers of events in data, results of fits for sample composition and \afbl, 
         and predictions for \afbl. The asymmetries are given at reconstruction level,
         with their total uncertainties. The sample compositions are given with 
         the fit uncertainties.
  \label{tab:afbl}
}
  \begin{tabular}{| l | c | c | c | c | c |}
    \hline
    & \multicolumn{1}{|c|}{$l$+$\ge$4 jets} & \multicolumn{1}{|c|}{$e$+$\ge$4 jets} & 
      \multicolumn{1}{|c|}{$\mu$+$\ge$4 jets} & \multicolumn{1}{|c|}{$l$+4 jets} & \multicolumn{1}{|c|}{$l$+$\ge$5 jets} \\
    \hline 
    Raw \Nlfwd & \multicolumn{1}{|c|}{867} & \multicolumn{1}{|c|}{485} & 
      \multicolumn{1}{|c|}{382} & \multicolumn{1}{|c|}{730} & \multicolumn{1}{|c|}{137}\\
    Raw \Nlbwd & \multicolumn{1}{|c|}{665} & \multicolumn{1}{|c|}{367} & 
      \multicolumn{1}{|c|}{298} & \multicolumn{1}{|c|}{546} & \multicolumn{1}{|c|}{119} \\[2ex]
    \Ntt        & $1096 \pm 39$ & $622\pm28$ & $474\pm27$ & $881\pm36$ & $211\pm16$\\
    \Nw         &  $356\pm39$ & $173\pm28$ & $198\pm27$ & $323\pm36$ &  $31\pm16$\\
    \Nmj        &   $79\pm 5$ &  $56\pm 3$ &   $8\pm 2$ &  $66\pm 4$ &  $14\pm2$\\
    \afbl\ (\%) &   $14.2\pm 3.8$ &  $16.5\pm 4.9$ &  $9.8\pm 5.9$ &  $15.9\pm 4.3$ &   $7.0\pm 8.0$\\[2ex]
    \mcatnlo\ \afbl\ (\%) &   $0.8\pm 0.6$ &  $0.7\pm 0.6$ &  $1.0\pm 0.8$ &  $2.1\pm0.6$ &   $-3.8\pm1.2$\\
    \hline
  \end{tabular}
\end{table}

\begin{table}[htbp]
\caption{
  Reconstruction-level \afb\ by subsample.
  \label{tab:subsamples}
}

\begin{tabular}{|l|c|c|}
\hline
  &  \multicolumn{2}{|c|}{\afb\ (\%)} \\
\hline
Subsample  & \multicolumn{1}{|c|}{Data} & \multicolumn{1}{|c|}{\mcatnlo} \\ 
\hline 
$\mttbar<450\GeV$ &  $7.8 \pm 4.8$ & $1.3 \pm 0.6$  \\   
$\mttbar>450\GeV$ & $11.5 \pm 6.0$ & $4.3 \pm 1.3$  \\
$\absdy<1.0$      &  $6.1 \pm 4.1$ & $1.4 \pm 0.6$  \\
$\absdy>1.0$      & $21.3 \pm 9.7$ & $6.3 \pm 1.6$ \\
\hline
\end{tabular}
\end{table}

\begin{figure}[ht]
\centering
\includegraphics[width=70mm]{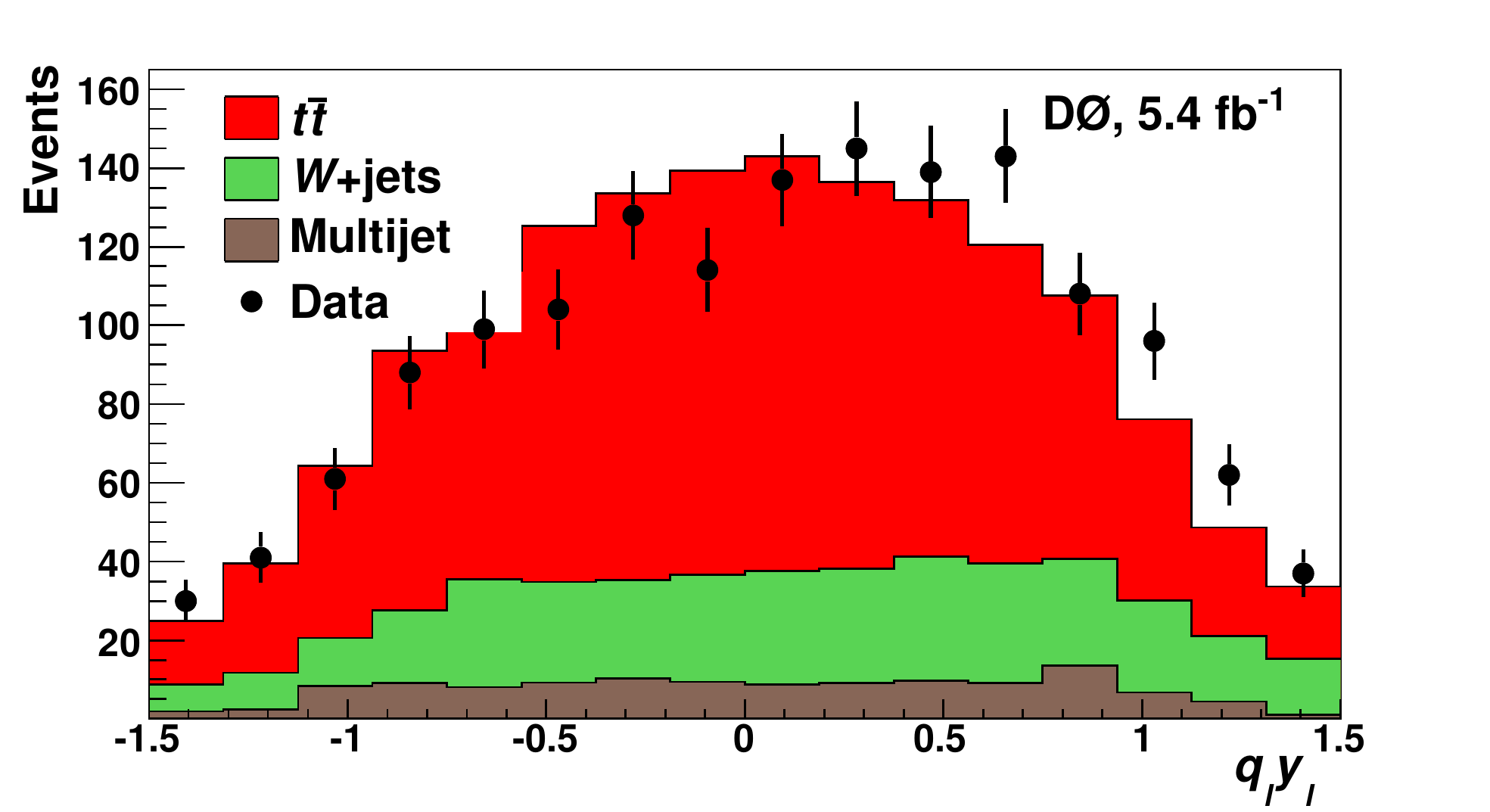}
\includegraphics[width=70mm]{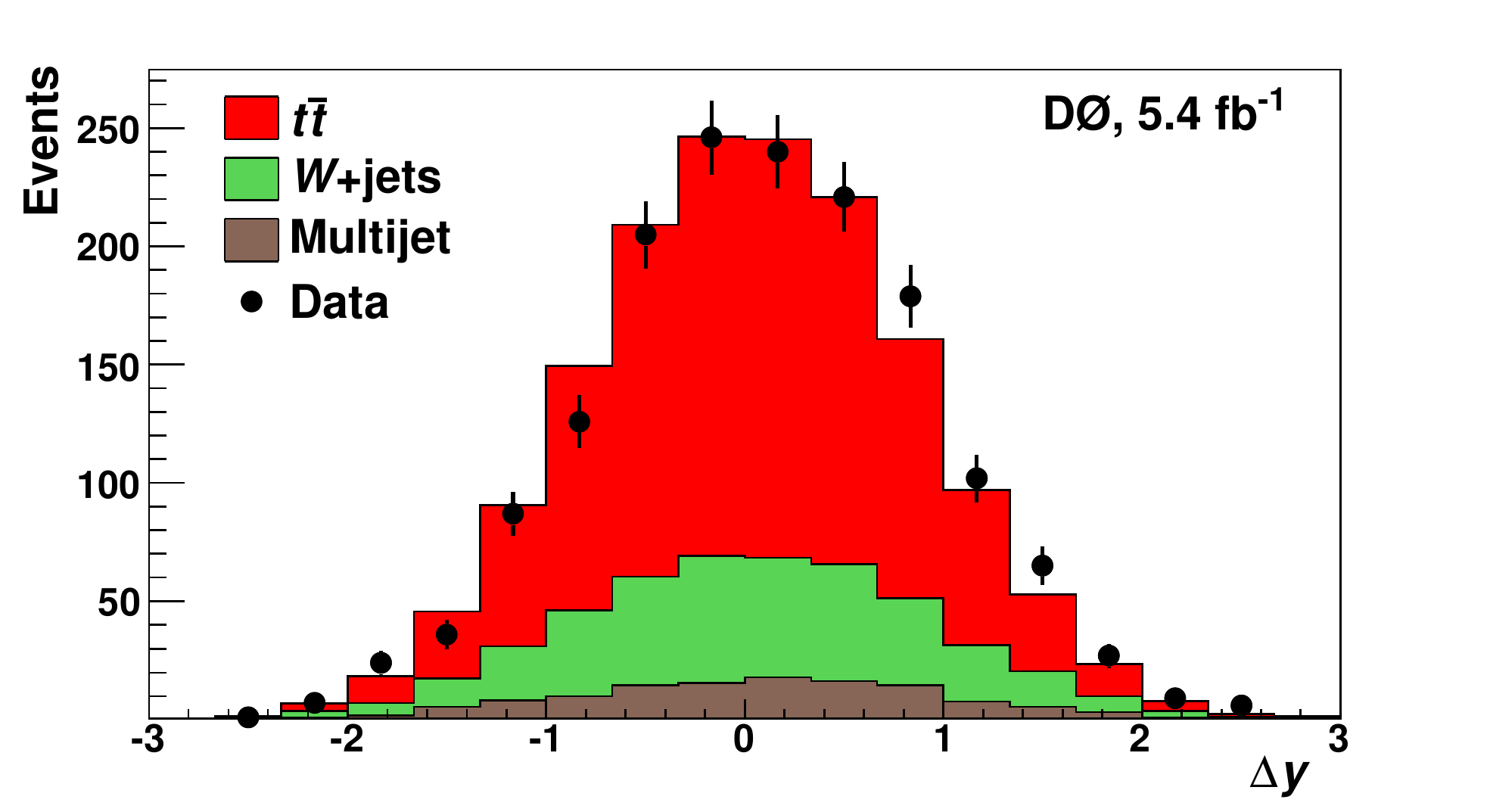}
\begin{picture}(10,10)
\put(-250,80){(a)}
\put(-50,80){(b)}
\end{picture}
\caption{Distribution in \qyl\ (a) and \dy\ (b) for data and simulation. The sample composition is taken from the fit.} \label{fig_reco}
\end{figure}

\section{Unfolding Technique and Results}

To infer the asymmetry at the production level, we use regularized unfolding to correct the reconstruction level \dy\ distribution, which is the \dy\ distribution in data after subtracting the contribution from the backgrounds. Unfolding corrects for effects from detector reconstruction and acceptance, which change \dy\ at the reconstruction level. In order to take into account the fine details of the reconstruction the unfolding procedure, based on a modified version of ROOT's \rm{TUnfold} class, uses a migration matrix with 50 bins for the reconstruction level and 26 bins for the production level~\cite{bib:tunfold}. The acceptance correction multiples each bin by $\frac{1}{eff_{i}}$, where $eff_{i}$ is the selection efficiency for bin $i$ in \dy. Regularization smoothes the large statistical fluctuations between bins in the \dy\ distribution. The unfolded $\dy$ distribution is summarized in the two bins of the asymmetry.

In addition to regularized unfolding, we also perform a four-bin maximum likelihood unfolding of the \dy\ distribution. For the four-bin procedure, the boundaries of the bins are at $\dy = -3, -1, 0, 1, 3$. The result of the four-bin unfolding procedure is $\afb = (16.9 \pm 8.1)\%$.

Because migrations in between bins for the lepton-based are very small, only an acceptance correction is used to unfold the \qyl\ distribution. Results from the unfolding are shown in Table~\ref{tab:dyafb}.

\begin{figure}[ht]
\centering
\includegraphics[width=70mm]{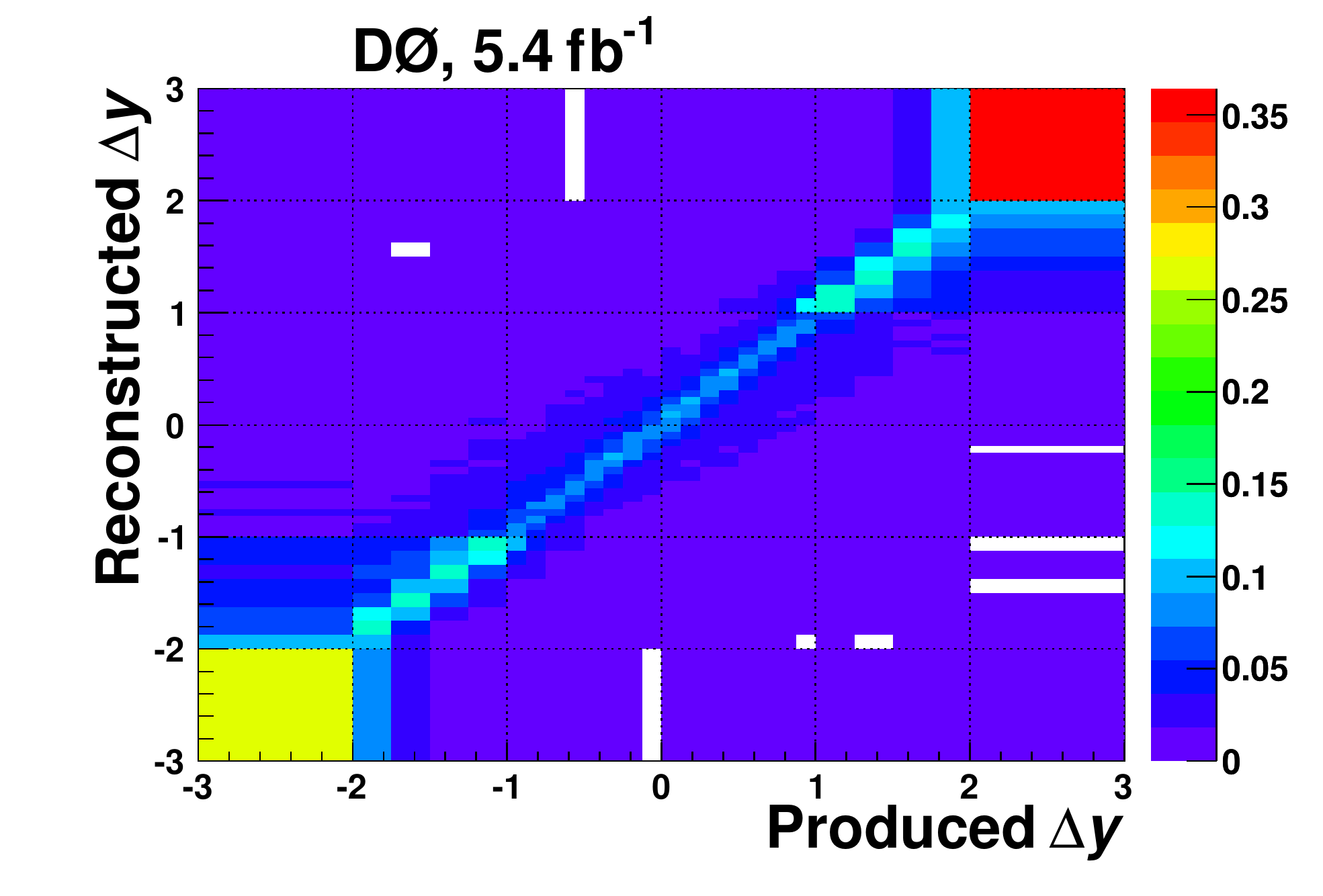}
\caption{The 50 by 26 bin migration matrix used for the regularized unfolding.} \label{fig_mig}
\end{figure}

\begin{table}[htbp]
\caption{
  \dy- and lepton-based Asymmetries. 
  \label{tab:dyafb}
}

\begin{tabular}{| l | c | c | c | c |}
\hline
  &  \multicolumn{2}{|c|}{\afb\ (\%)} & \multicolumn{2}{|c|}{\afbl\ (\%)} \\  
  \hline
  & \multicolumn{1}{|c|}{Reconstruction level} & \multicolumn{1}{|c|}{Production level}  & \multicolumn{1}{|c|}{Reconstruction level} & \multicolumn{1}{|c|}{Production level} \\ 
\hline
Data     & 9.2$\pm$3.7 & 19.6$\pm$6.5 & 14.2 $\pm$ 3.8 & 15.2 $\pm$ 4.0 \\
\mcatnlo & 2.4$\pm$0.7 &  5.0$\pm$0.1 & 0.8 $\pm$ 0.6&  2.1 $\pm$ 0.1 \\
\hline
\end{tabular}
\end{table}

\section{Systematic Uncertainties}

The systematic uncertainties, shown in Tables~\ref{tab:uncert} and~\ref{tab:uncert_lep}, are computed for both the reconstruction and production level measurements, as well as for the reconstruction level predictions. Note that although we make a detailed accounting of the systematics effects, the measurement is dominated by statistical uncertainty. The uncertainties into seven different categories. Further details about each category of uncertainty may be found in Ref~\cite{Abazov:2011rq}.

\begin{table}

\caption{Absolute uncertainty on \afb\ (\%)}
\begin{tabular}{ | l | c | c | c |}
\hline

\hline
       &  \multicolumn{2}{|c|}{Reco. level} & Prod. level \\
\hline
Source & Prediction & Measurement & Measurement\\
\hline
Jet reco   &    $\pm 0.3$        &        $\pm 0.5$         &   $\pm 1.0$      \\
JES/JER   & $+0.5$            &      $-0.5$          &     $-1.3$     \\
Signal modeling &  $\pm 0.3$          &    $\pm 0.5$            &  ${+0.3}/{-1.6}$        \\
$b$-tagging   & \emph{-}            &    $\pm 0.1$       &   $\pm 0.1$   \\
Charge ID  &    \emph{-}  & $+0.1$ & ${+0.2}/{-0.1}$     \\
Bg subtraction &  \emph{-} &  $\pm0.1$ & ${+0.8}/{-0.7}$ \\
Unfolding Bias   &   \emph{-}            &  \emph{-}    & ${+1.1}/{-1.0}$\\ 
\hline
Total                            & ${+0.7}/{-0.5}$ &  ${+0.8}/{-0.9}$ & $+1.8/{-2.6}$\\
\hline
\end{tabular}
\label{tab:uncert}
\end{table}

\begin{table}
\caption{Absolute uncertainty on \afbl\ (\%)}
\begin{tabular}{| l | c | c | c |}
\hline
       &  \multicolumn{2}{|c|}{Reco. level} & Prod. level \\
\hline
Source & Prediction & Measurement & Measurement\\
\hline 
Jet reco &    $\pm 0.3$        &      $\pm 0.1$       &  $\pm 0.8$      \\
JES/JER  & $+0.1$            &    $-0.4$          &      ${+0.1}/{-0.6}$     \\
Signal modeling   &  $\pm 0.3$          &    $\pm 0.5$           &   ${+0.2}/{-0.6}$        \\
$b$-tagging   &     \emph{-}        &     $\pm 0.1$    &    $\pm 0.1$   \\
Charge ID  &    \emph{-}  & $+0.1$ &  ${+0.2}/{-0.0}$     \\
Bg subtraction &  \emph{-} &  $\pm 0.3$ &  $\pm 0.6$ \\
\hline
Total                            & $\pm 0.5$ & $\pm0.7$ & ${+1.0}/{-1.3}$\\
\hline
\end{tabular}
\label{tab:uncert_lep}
\end{table}

\section{Cross checks}

We perform multiple cross checks to ensure that the measurement is accurate. The asymmetry from the \wpj\ background is taken from simulation, specifically \alpgen+\pythia. To check that the simulated asymmetry is accurate, we use a similar template-based measurement which includes events without a $b$-tagged jet, as seen in Figure~\ref{fig_xcheck}, but meeting every other event selection criteria. These events are dominated by the \wpj\ background.  The fitted asymmetry for \wpj\ using this method is in good agreement with the simulated asymmetry.

The polarities of the solenoid and toroid magnets of the \DZ\ detector are regularly flipped. To ensure that there is no inherent bias in the asymmetry from the detector, measurements are made for each of the four polarity settings for both the \dy- and lepton-based asymmetries. No significant differences were found between samples with different settings of magnet polarities. Similarly, breaking the lepton-based asymmetry into samples with positive and negative charge did not significantly affect the result. 

\begin{figure}[ht]
\centering
\includegraphics[width=70mm]{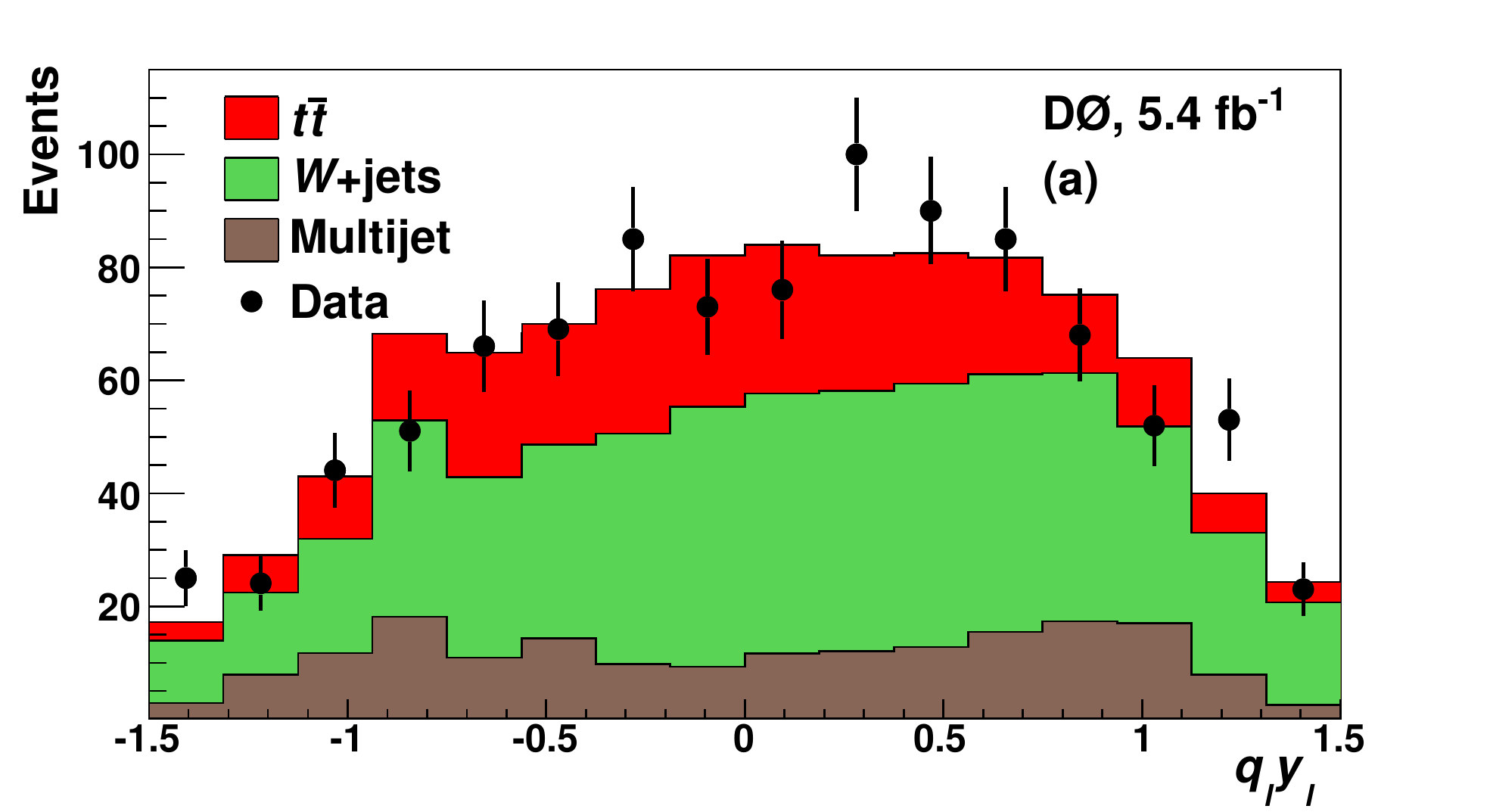}
\includegraphics[width=70mm]{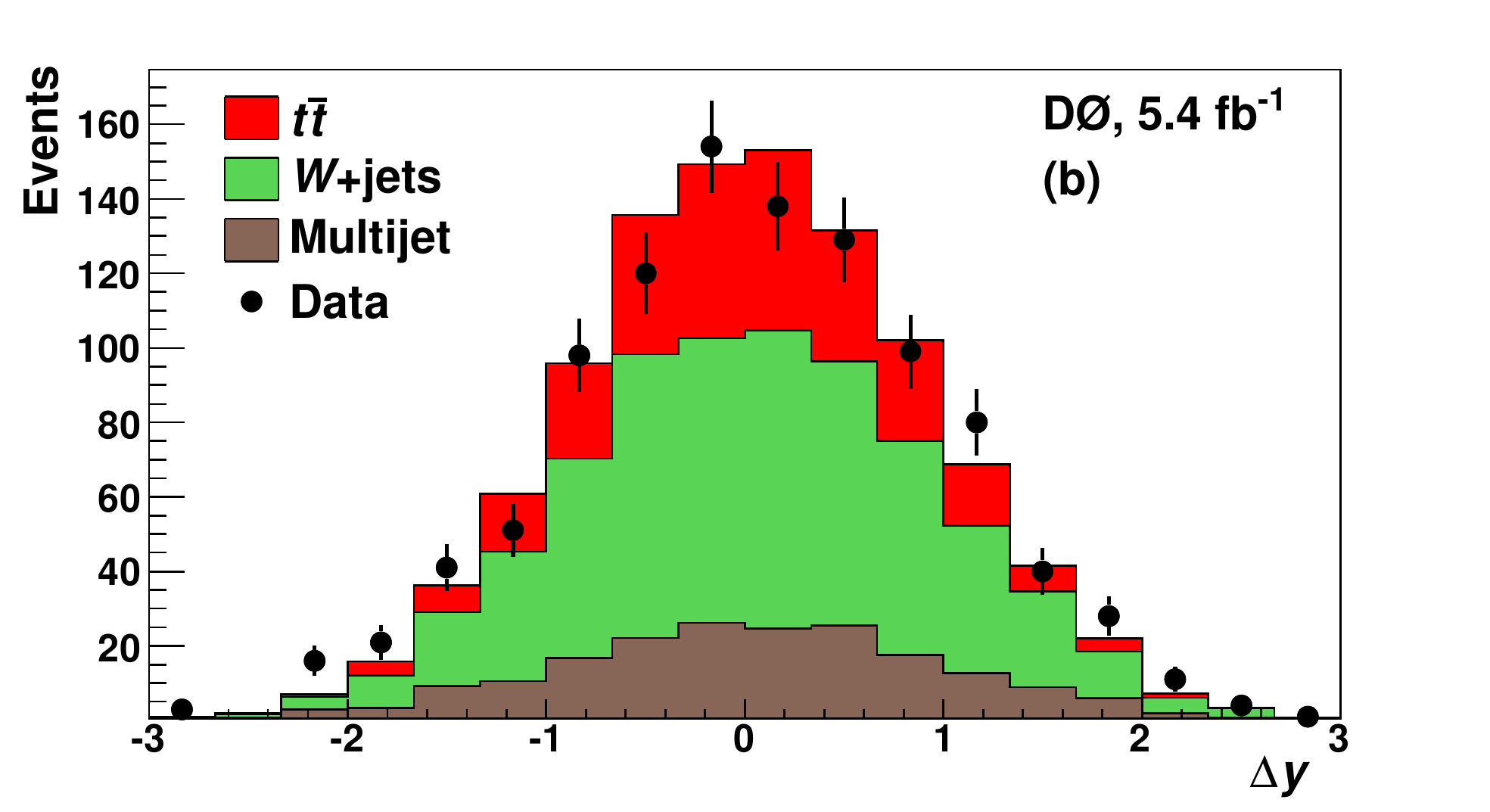}
\caption{Distributions in \qyl\ (a) and \dy\ (b) for events with zero $b$-tagged jets. The sample composition is taken from the fit.} \label{fig_xcheck}
\end{figure}

\section{Discussion}

In summary, the reconstruction level results for asymmetries based on \dy\ and \qyl\ are presented in Tables~\ref{tab:afbcomb} and~\ref{tab:afbl}, respectively. Results from unfolding to compare to the production level predictions are shown in Table~\ref{tab:dyafb}. The measured values for both asymmetries are higher than the predictions \mcatnlo.

In addition to the measurements, we investigated the behavior of asymmetry on gluon radiation via the transverse momentum of the \ttbar\ system, \ttpt. In the SM, the asymmetry arises at next-to-leading-order from an interference of various diagrams. Diagrams which contain external gluon lines, either in the initial state or the final state, interfere to produce a negative asymmetry. On the other hand, the interference between diagrams in which all gluon lines are internal produces a positive asymmetry.

Because \ttpt\ is made up of the six detected objects from \ttbar\ decay, the recoil from low energy gluon radiation can be measured. The dependence of the asymmetry across the \ttpt\ spectra contains both soft and hard gluon radiation, and the behavior is shown in fine detail in Figure~\ref{fig_dep}. The dependence of \afb\ on \ttpt\ simulated with \mcatnlo\ and \pythia\ is shown. Two different dependences are shown for \pythia, one with the angular coherence (angular ordering) parameter on (shown for the curve corresponding to \pythia\ 6.425 D6-Pro). The other curve, for the \pythia\ 6.425 S0A-Pro setting, has angular coherence turned off, and the dependence of \afb\ on \ttpt\ is no longer present. We note that we only use \pythia\ for qualitative purposes, and \mcatnlo\ is used for all predictions. Because the selection efficiency versus \ttpt\ is not constant (Figure~\ref{fig_dep}), we turn off the dependence of \afb\ and \ttpt\ for one of the systematics.

We briefly mention the comparison of data and simulation for \ttpt. In Figure~\ref{fig_ttpt}, we show the distribution in \ttpt\ for both \mcatnlo\ and \pythia\ with initial state radiation (ISR) turned off. Here \pythia\ with modified initial state radiation does a better job describing the data. Two things should be noted. One, the resolution for \ttpt\ is poor. Bins widths are shown with half the resolution, meaning two bins is about the best it can get. Two, the distribution in the number of jets, which agrees very well with data using \mcatnlo, but no longer agrees while when using \pythia\ with ISR turned off. We present no concrete conclusion for this issue, but show what we have found so far.

\begin{figure}[ht]
\centering
\includegraphics[width=80mm]{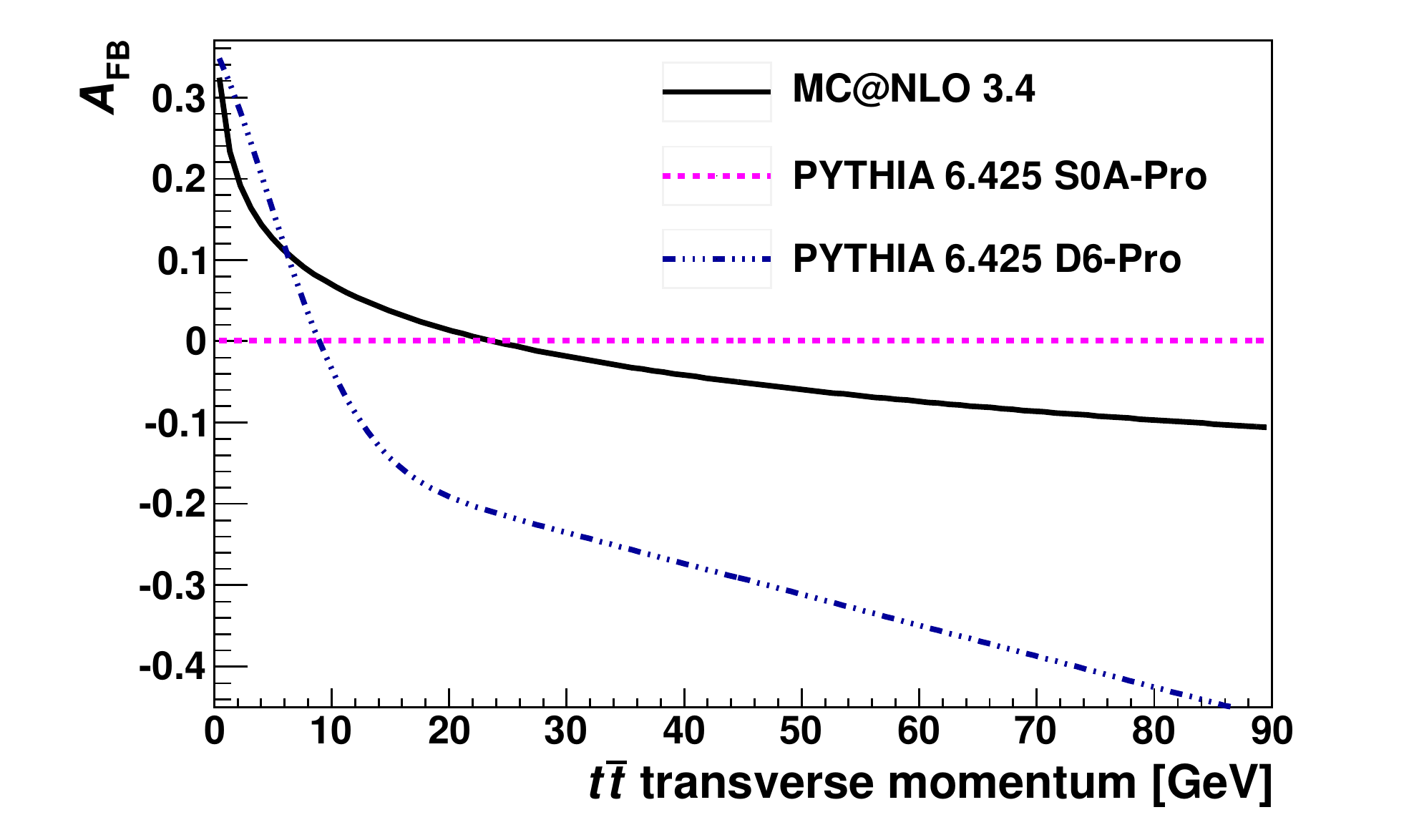}
\includegraphics[width=71mm]{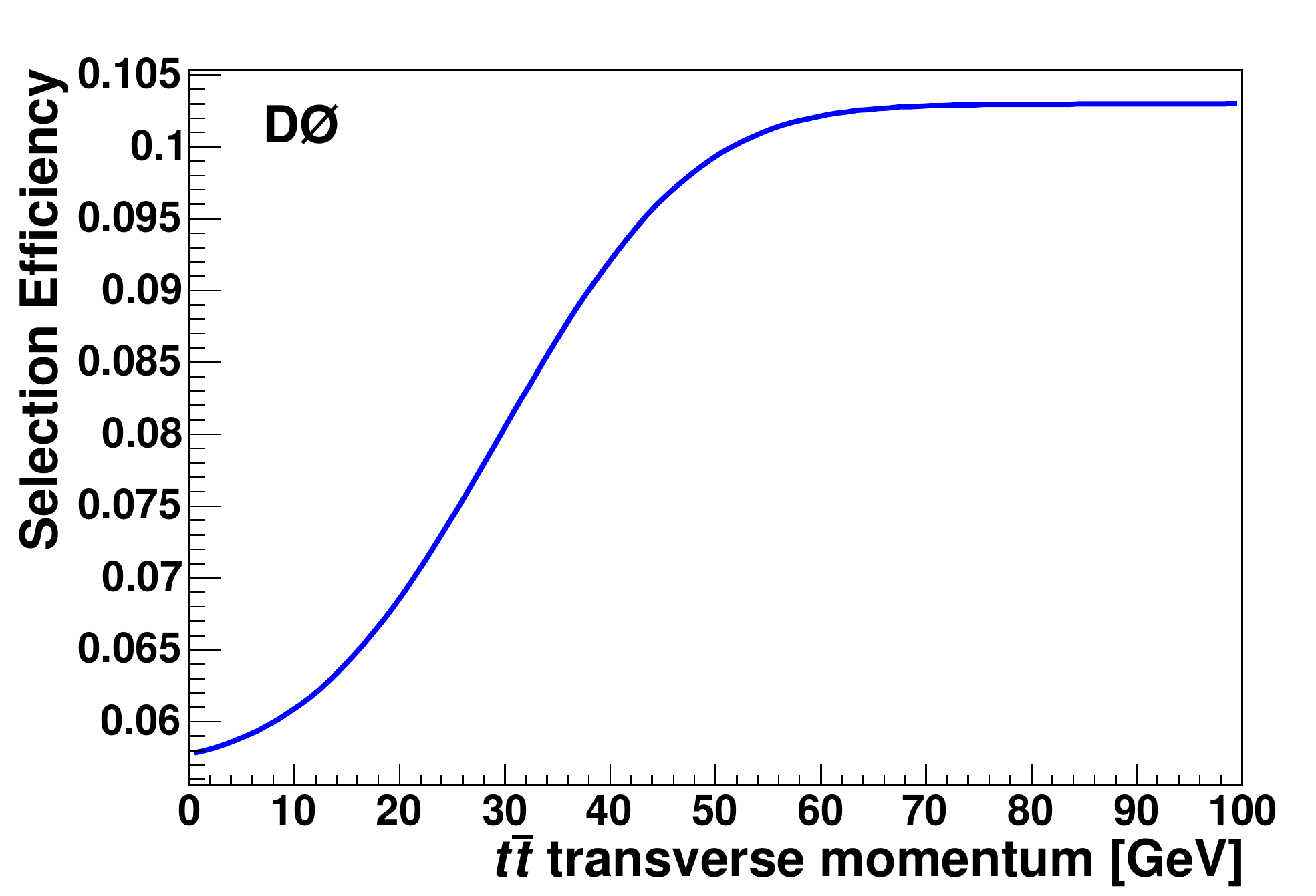}
\begin{picture}(10,10)
\put(-375,100){(a)}
\put(-150,100){(b)}
\end{picture}
\caption{\afb\ versus \ttpt\ for \mcatnlo\ (black) and \pythia\ with different settings (a). The selection efficiency versus \ttpt\ (b). } \label{fig_dep}
\end{figure}

\begin{figure}[ht]
\centering
\includegraphics[width=70mm]{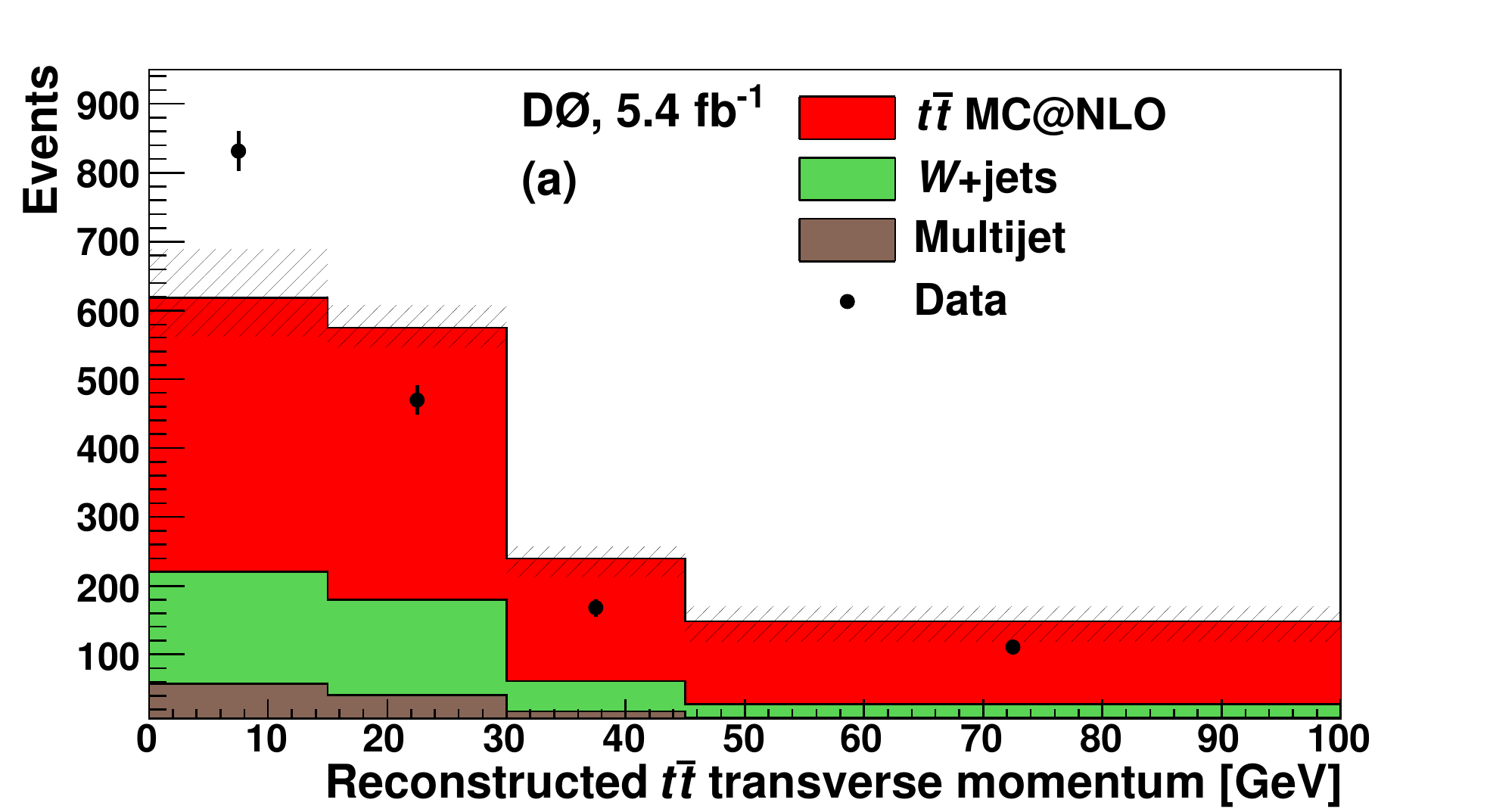}
\includegraphics[width=70mm]{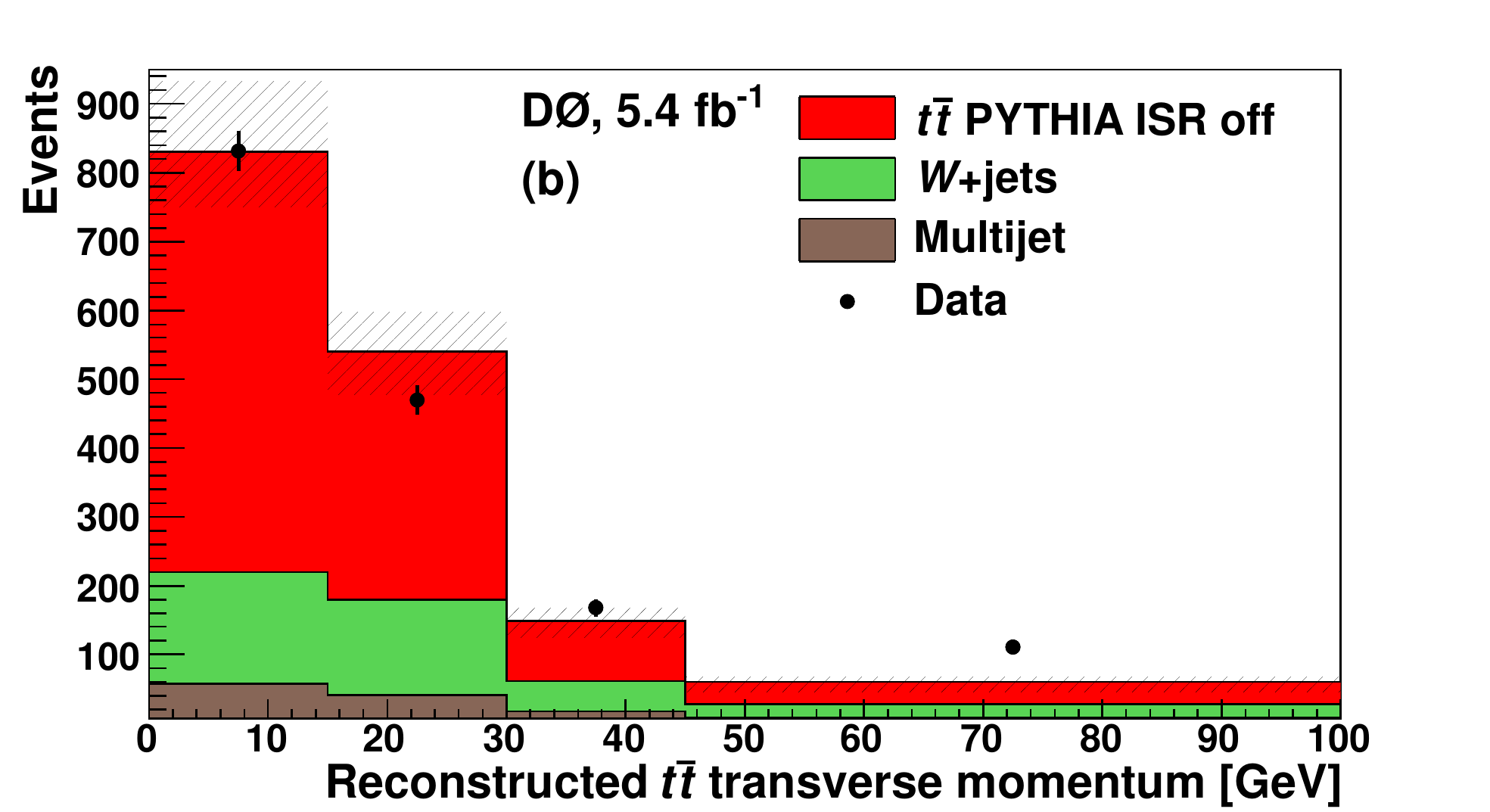}
\caption{Distribution in \ttpt\ for data and simulation for \mcatnlo\ (a) and \pythia\ with initial state radiation turned off (b).} \label{fig_ttpt}
\end{figure}

\section{Conclusion}

We present measurements of the forward-backward asymmetry in \ttbar\ events using both \dy\ and \qyl. At the reconstruction level, the \dy-based asymmetry is (9.2 $\pm$ 3.7)\% in data and (2.4 $\pm$ 0.7)\% for the \mcatnlo-based prediction. The lepton-based reconstruction level asymmetry is (14.2 $\pm$ 3.8)\% for data and (0.8 $\pm$ 0.6)\% for the \mcatnlo-based prediction.

In addition to the reconstruction level measurements, we also present unfolded asymmetries for \dy\ and \qyl. Unfolding the \dy\ distribution gives an asymmetry of (19.6 $\pm$ 6.5)\%, compared to \mcatnlo\ production level predictions of (5.0 $\pm$ 0.1)\%. In comparison, unfolding the \qyl\ distribution results in an asymmetry of (15.2 $\pm$ 4.0)\%, to be compared with (2.1 $\pm$ 0.1)\%. We note that other SM predictions measure larger asymmetries. 

The measurements from \DZ\ data are consistently higher than the predictions. We mentioned some potential limitations of these predictions.

\begin{acknowledgments}
The author would like to thank to K. Melnikov and S. Mrenna for enlightening discussions.
\end{acknowledgments}

\bigskip 

\end{document}